\documentclass[a4paper]{llncs}

\usepackage[english]{babel}
\usepackage[utf8x]{inputenc}
\usepackage[T1]{fontenc}

\usepackage{amsmath}
\usepackage{graphicx}
\usepackage[colorinlistoftodos]{todonotes}
\usepackage[colorlinks=true, allcolors=blue]{hyperref}

\newcommand{\Vfire}{V_T} 

\title{Fractal analyses of networks of integrate-and-fire stochastic spiking neurons}

\author{Ariadne A. Costa\inst{1}, Mary Jean Amon\inst{1}, Olaf Sporns\inst{1}, Luis H. Favela\inst{2}}

\institute{Department of Psychological and Brain Sciences, Indiana University, Bloomington-IN, 47405, USA\\
\and
Department of Philosophy and Cognitive Sciences Program, University of Central Florida, Orlando-FL, 32816, USA}

\begin{document}
\maketitle

\begin{abstract}
Although there is increasing evidence of criticality in the brain, the processes that guide neuronal networks to reach or maintain criticality remain unclear. The present research examines the role of neuronal gain plasticity in time-series of simulated neuronal networks composed of integrate-and-fire stochastic spiking neurons, and the utility of fractal methods in assessing network criticality. Simulated time-series were derived from a network model of fully connected discrete-time stochastic excitable neurons. Monofractal and multifractal analyses were applied to neuronal gain time-series. Fractal scaling was greatest in networks with a mid-range of neuronal plasticity, versus extremely high or low levels of plasticity. Peak fractal scaling corresponded closely to additional indices of criticality, including average branching ratio. Networks exhibited multifractal structure, or multiple scaling relationships. Multifractal spectra around peak criticality exhibited elongated right tails, suggesting that the fractal structure is relatively insensitive to high-amplitude local fluctuations. Networks near critical states exhibited mid-range multifractal spectra width and tail length, which is consistent with literature suggesting that networks poised at quasi-critical states must be stable enough to maintain organization but unstable enough to be adaptable. Lastly, fractal analyses may offer additional information about critical state dynamics of networks by indicating scales of influence as networks approach critical states.
\end{abstract}

\section{Introduction}
The last two decades have seen increasing discussion about the prevalence and role of criticality in the brain~\cite{Chialvo2004,Beggs2012,Favela2014,Hesse2014}. Criticality is a property of systems organized near phase transitions. Loosely speaking, a critical state is stable enough to maintain organization, but is unstable enough to be adaptable so as to facilitate switches among 
states. Criticality has been experimentally demonstrated in neuronal systems in a variety of ways, for example, \textit{in vivo}~\cite{Poil2008,Petermann2009,Hahn2010,Favela2016}, \textit{in vitro}~\cite{Beggs2003,Beggs2004} and~\textit{in silico} (e.g., \cite{Arcangelis2006,Levina2007,Costa2015,Brochini2016}). A natural question to ask is, “Why do brains exhibit criticality?” Though there is no agreed upon answer, a number of possibilities have been offered; for example, criticality maximizes the range of inputs that can be processed by neurons~\cite{Kinouchi2006,Shew2009}, optimizes information processing~\cite{Beggs2008,Shew2013}, and is a signature of brain dynamics in healthy nervous systems~\cite{Massobrio2015}.

A primary indicator of criticality is power-law distributions. As an example, both size and duration of neuronal avalanches -- cascades of neurons spiking consecutively -- in cortical circuits are distributed according to power laws~\cite{Beggs2003}. In regard to criticality, it is claimed that power-law distributions exhibit fractal scaling. Fractals are scale-free and self-similar spatial or temporal patterns, whereby the global pattern is maintained at various scales of observation~\cite{Mandelbrot1982,Bak1987}. Mathematical fractals are perfectly self-similar across scale (e.g., Koch snowflakes and Sierpinski triangles). Natural fractals are statistically self-similar across scales (e.g., the coastline of Britain and tree branching). In this sense, power-law distributions can represent the scale-free self-similarity of fractals, although precise relationship between criticality, fractals, and power laws remains controversial (for review see~\cite{Watkins2016}). While there is growing evidence of criticality in brains, an explanation of how brains reach and maintain such states remains elusive (cf. \cite{Tetzlaff2010}).

Considering that power-law distributions can be indicators of criticality, and given that fractal scaling can be represented in terms of power-law distributions, fractal analyses may be useful methods for assessing the presence of criticality and quantifying how far a system is from criticality. In the current work, we test this hypothesis and attempt to understand more about the processes that guide a neuronal network to reach and/or maintain itself around criticality. To do this, we determine the scaling behaviors of temporal series of simulations of neuronal networks composed of stochastic spiking neurons~\cite{Galves2013} with gain plasticity~\cite{Costa2017} using monofractal detrended fluctuation analysis (DFA) and multifractal DFA (MFDFA). This neuronal model was first proposed as an explanation for self-organizing criticality (SOC) in cortical circuits in the brain, suggesting that neural circuits operate slightly above criticality (self-organizing supercriticality; SOSC).

In a classic paper, Bak, Tang and Wiesenfeld showed that systems with spatial degrees of freedom self-organize into critical states reflected by $1/f$ (fractal) noise spectra, presenting minimally stable clusters at all length scales~\cite{Bak1987}. In accordance with Bak and colleagues, our main finding is that $1/f$ noise in time series of neuronal gain plasticity corresponds to systems in (quasi-)critical states.

In the following section, we detail the mathematical neuronal model implemented and also the methods used for the analyses. Next, we present the fractal analysis results, and conclude with a discussion of the results that may further illuminate the nature of neuronal criticality.

\section{Method}

\subsection{Neuronal model}
\label{section:neuronal_model}
The model considered here was first proposed by Costa \textit{et al.}~\cite{Costa2017}. It consists of a network of $i=1,\ldots,N$ discrete-time stochastic excitable neurons~\cite{Gerstner1992,Gerstner2002,Galves2013,Brochini2016,Costa2017}. Each neuron is connected to all other neurons $j$ (i.e., it is a fully connected network). The presynaptic neuron $j$ transmits signals to the postsynaptic neuron $i$ proportionally to the synaptic strength $W_{ij}$.

This is an adaptation of the Galves-Löcherbach (GL) model~\cite{Galves2013} by adding neuronal gain plasticity. As in the GL model, each neuron has a membrane potential $V_i$ that is evolved at each timestep $t$. In the special case of GL model with the filter function $g(t-t_s) = \mu^{t - t_s}$, 
where $t_s$ is the time of the last firing of neuron $i$, the membrane potential at $t+1$ can be discretized as:

\begin{equation}
V_i[t+1] = 
  \left\{
     \begin{array}{lcl}
        \displaystyle
        0  &\quad& \hbox{if $X_i[t]= 1$,} \\
        \displaystyle
         \mu V_i[t] + I_{ext} + \frac{1}{N}
         \sum_{j=1}^{N} W_{ij} X_j[t] &\quad&
            \hbox{if $X_i[t] = 0$,}
     \end{array}
   \right.
\label{eq.Potential}
\end{equation}

\noindent where $I_{ext}$ represents external stimuli arriving at the postsynaptic neuron, while $X_j$ is the state of the presynaptic neuron between the timesteps $t$ and $t+1$ ($X_j = 1$ when $j$ spikes and $X_j = 0$ otherwise). The leakage factor $\mu \in [0,1]$ reflects the diffusion of ions through the membrane. 

Unlike the classic leaky integrate-and-fire (LIF) model~\cite{Lapicque1907}, the neuron does not fire deterministically when $V_i[t+1]$ exceeds a threshold. That fixed threshold potential is substituted by a probability of firing $\Phi(V_i[t])$, according to a \textit{firing function}~\cite{Galves2013,Larremore2014,Duarte2014,DeMasi2015,Galves2016,Costa2017}. The \textit{rational} firing function~\cite{Larremore2014,Brochini2016,Costa2017} is used to calculate the firing probability $0 \leq ~ \Phi(V_i[t]) \leq 1$ for each neuron at each timestep:

\begin{equation}
 \Phi_i[t](V_i[t]) = \frac{\Gamma_i[t] (V_i[t]-\Vfire)}{1+\Gamma_i[t]-(V_i[t]-\Vfire)} \:
  \Theta(V_i[t]-\Vfire)\:. 
  \label{eq.Rational}
\end{equation}

The neuronal gain $\Gamma_i$ causes an amplification to the signal received by the neuron, and is evolved as it follows:

\begin{equation}
 \Gamma_i[t+1] = \Gamma_i[t] +\frac{1}{\tau} 
 \Gamma_i[t] - \Gamma_i[t] X_i[t]\:,
\label{eq.Gain}
\end{equation} 

\noindent where $\tau$ is gain recovery time. The neuronal gain recovers each timestep and is decreased just after the neuron spikes. These dynamics are biologically plausible, corresponding to the reduction and recovery of sodium channels at the axon initial segment after spiking, as described in~\cite{Kole2012}. 

An advantage of studying neuronal gain plasticity instead of synaptic
plasticity —-- as in previous works~\cite{Levina2007,Costa2015,Campos2017} --- is the reduction in the number of equations evolved at each timestep: $N$ equations for neuronal gains rather than $N(N-1)$ equations for corresponding synapses~\cite{Brochini2016,Costa2017}. 

The results presented here come from simulations of networks with $N=160,000$ neurons without external stimulus ($I_{ext}=0$). After spiking, neurons do not have memory of previous timesteps ($\mu=0$). The threshold potential is $\Vfire=0$. Simulations were conducted in Fortran90. 

Analyses were performed in time series of average gains (computed over all neurons at each timestep) for different values of neuronal gain recovery times ($\tau$). A long transient of 5 million
timesteps was removed (see~\cite{Costa2017}), so that only the last 50,000 timesteps were analyzed.

\subsection{Fractal analyses}
Detrended fluctuation analysis (DFA) is a type of monofractal analysis that removes local linear trends within specified windows of time in the data, and then looks for statistical self-similarity in what remains. After linear detrending, the residual represents fluctuations around the global trend. For each window size the log-log plot of the transformed frequency as a function of the transformed amplitude fluctuations reveals a linear relation indicating the degree of self-similarity across scaling, given by the Hurst exponent ($H$). 

Hurst exponents approaching one ($H \approx 1$) represent $1/f$ noise or $1/f$ scaling. $1/f$ scaling indicates the presence of fractal structure within a signal, or self-similar temporal or spatial patterns across scales~\cite{Kello2007,Holden2009,VanOrden2011}. $1/f$ noise contrasts with white noise, which represents relatively random or independent timesteps ($H \approx 0.5$). Hurst exponents close to 1.5 or higher ($H \approx 1.5$) represent Brownian motion. Brownian noise describes patterns of variability that exhibit a random walk pattern, with global structure and local independence~\cite{Gilden2001,Holden2005}. Brownian noise often can be used to describe the movement of natural systems, where it is not easy to predict a specific movement trajectory but the trajectory is always dependent on the system’s previous position. Lastly, blue noise is indicative of anti-persistence ($H \approx 0.0$), where positive data points tend to be followed by negative ones and vice versa such that this signal tends toward its mean. Anti-persistence is an indicator of a signal with short memory~\cite{Delignieres2006,Holden2005}. 

Monofractal scaling identifies one scaling relationship that best characterizes a signal and assumes invariance across temporal or spatial scales. However, variance often occurs across scales. Multifractal analysis indicates the degree to which power-law structure and self-similarity across scales is heterogeneous across a signal. Unlike monofractal analysis, multifractal analysis is capable of characterizing different local scaling properties across a signal~\cite{Ihlen2012}. Multifractal detrended fluctuation analysis (MFDFA) adds an additional $q$ parameter to DFA, which weights the influence of small and large fluctuations or root-mean-square (RMS). RMS of variation around local trends is successively raised to the value of each $q$ parameter~\cite{Favela2016}. The more negative the $q$ value, the more strongly it is influenced by segments with small RMS. Conversely, more positive $q$ value are influenced by segments with large RMS and $q$’s of 0 are neutral to the influence of relatively small or large RMS~\cite{Ihlen2012}. The variation of Hurst exponents ($H_{max}$ - $H_{min}$) based on $q$ provides an index of multifractality, or the degree to which $1/f$ scaling varies across the signal~\cite{Ihlen2012}. The multifractal spectrum can be plotted to represent the variation, as well as the relative length of the right and left tails of the multifractal spectra. Multifractal spectra with long right tails indicate that the fractal structure is relatively insensitive to local fluctuations with large magnitudes, and long left tails suggest that spectra are relatively insensitive to local fluctuations with small magnitudes~\cite{Ihlen2012}. 

Prior to analyses, data were normalized and outliers $\pm 4$ standard deviations were trimmed~\cite{Holden2005}. Minimum (min = 4) and maximum (max = 4,096) window sizes were selected to accommodate the sample size of the time-series ($S = 50,000$). The minimum window size was chosen to avoid local RMS fluctuation errors and a maximum window size was selected to represent a significant portion of the time series while allowing for multiple windows across the time series~\cite{Botcharova2014}. Windows were overlapped by 50\% to provide better estimates of within-window variability~\cite{Favela2016,Hardstone2012,Linkenkaer2001l}. A linear detrending procedure was utilized to examine power-law scaling in the residuals (e.g.~\cite{Botcharova2014}). Preliminary analyses indicated that data was fractional Brownian motion, suggesting that data did not need to be integrated prior to analysis~\cite{Eke2000,Delignieres2013}.

\section{Results}
\begin{figure}[!h]
 \centering
{\includegraphics[width=0.47\linewidth]{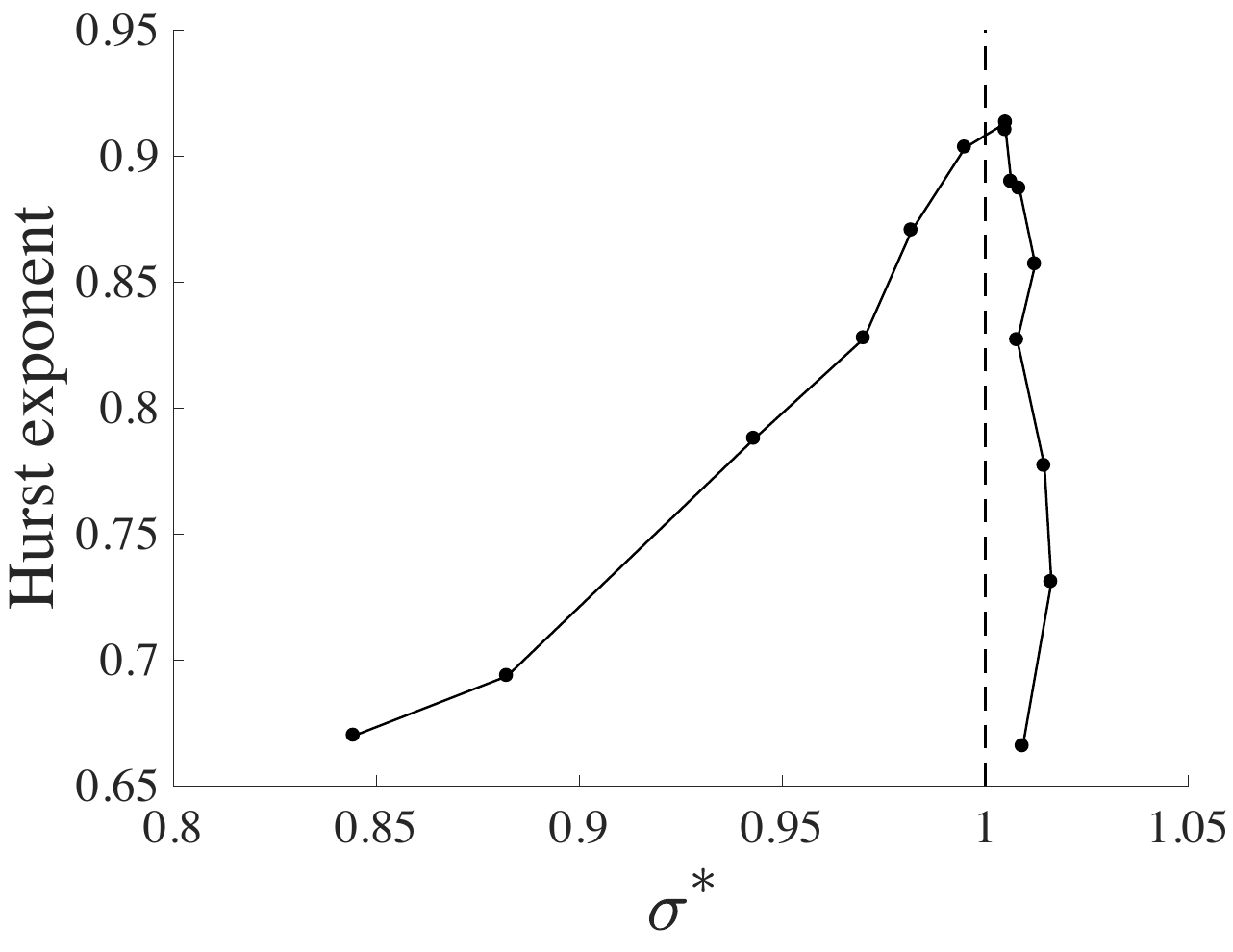}}\\(a)\\
{\includegraphics[width=0.47\linewidth]{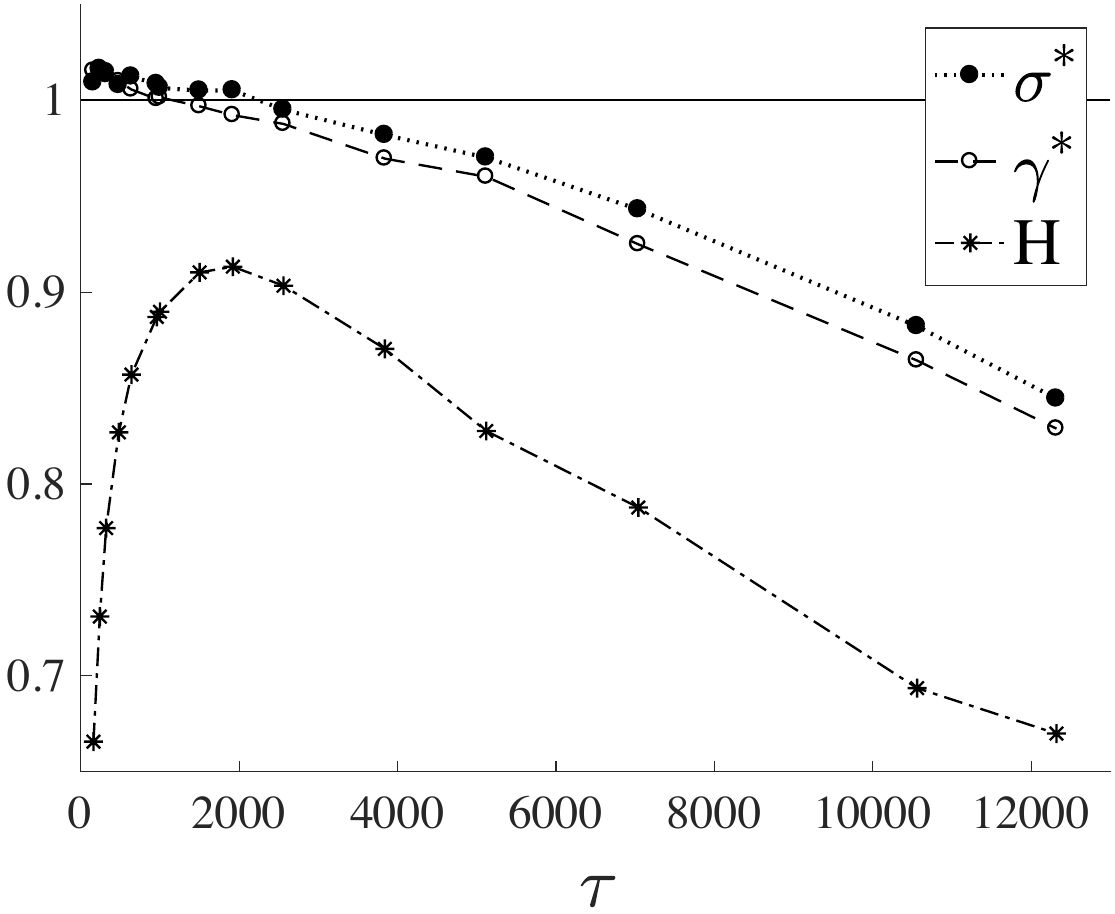}}\\ (b)
\caption{Relationship between characteristic neuronal gain recovery time ($\tau$) and fractal scaling (Hurst exponent), average branching ratio of active neurons ($\sigma^*$) and branching ratio. (a) As fractal scaling increases $\tau=160$ to 1,920 and peaks $\tau=1,920$, the branching ratio approaches criticality ($\sigma^* \approx 1$). After this point, the fractal scaling reduces, suggesting that greater fractal scaling of gain is associated with gain criticality. (b) As characteristic neuronal gain recovery time ($\tau$) increases, the branching ratio moves away from supercriticality ($\sigma^* >1$) to subcriticality ($\sigma^* <1$). The Hurst exponents peak for $\tau=1,920$.}
\label{fig:1}
\end{figure}

DFA was used to determine the extent to which a dynamical sequence of average neuronal gain ($\Gamma^*$) exhibited $1/f$ scaling. Hurst exponents were derived for average neuronal gain time-series for different $\tau$ values (Figure~\ref{fig:1}). Hurst exponents ($H$) peak in the range of $1/f$ (fractal) noise $\tau \approx 1,920$ ($H = 0.91$). Hurst exponents were positively associated with average gain ranging between $\tau = 160$ and $\tau = 1,920$. After peaking at $\tau \approx 1,920$, Hurst exponents began to decline with increasing $\tau$ values ($\tau = 1,920$ -- $12,320$). Hurst exponents were in the range between white and $1/f$ noise ($M = 0.81\pm 0.09$, min = 0.67, max = 0.91), with the greatest degree of $1/f$ scaling $\tau \approx 1,920$. 

For the same systems, the average branching ratio ($\sigma^*$) -- a common measure for criticality characterization~\cite{Haldeman2005,Costa2015,Timme2016,Wilting2016} -- was also computed (see Figure~\ref{fig:1}a). In theory, $\sigma^*=1$ corresponds to a stable (critical) system, while lower values of $\sigma^ *$ represent the activity decreasing in a subcritical system, and $\sigma^ *>1$ indicates the activity increasing in supercritical ones. Figure ~\ref{fig:1}b reveals how $H$ changes due to variations in the average branching ratio corresponding to the networks with different $\tau$ values. 

According to analytical calculations (see~\cite{Costa2017}), we can have the critical values $\Gamma_C = 1/W_C$ in the stationary state for the model with neither synaptic nor gain plasticity, i.e. fix $\Gamma$ and $W$ values for all neurons. In the simulations, we have fix $W_C=1$, which implies in $\Gamma_C=1$. However, for the model containing neuronal gain plasticity, as presented here, we can also assert due to analytical calculations that $\Gamma^*$ depends on $\tau$~\cite{Costa2017}. In this sense, we expect variations of the average gain in time series of networks with different $\tau$ values.

Figure \ref{fig:1}(b) compares three different measures to estimate criticality in neuronal networks with different characteristic times of recovery $\tau$. The three measures are the Hurst exponent ($H$), branching ratio ($\sigma^* $), and average gain ($\Gamma^* $). Peak $1/f$ noise, as well  $\Gamma^* \approx 1$ (for this specific case of the model with $\mu=0$, $V_T=0$, $I_{ext}=0$, $W_{ij}=W=1$) and $\sigma \approx 1$, are indicative of criticality.

\begin{figure}[!ht]
 \begin{center}  
  \includegraphics[width=0.71\linewidth]{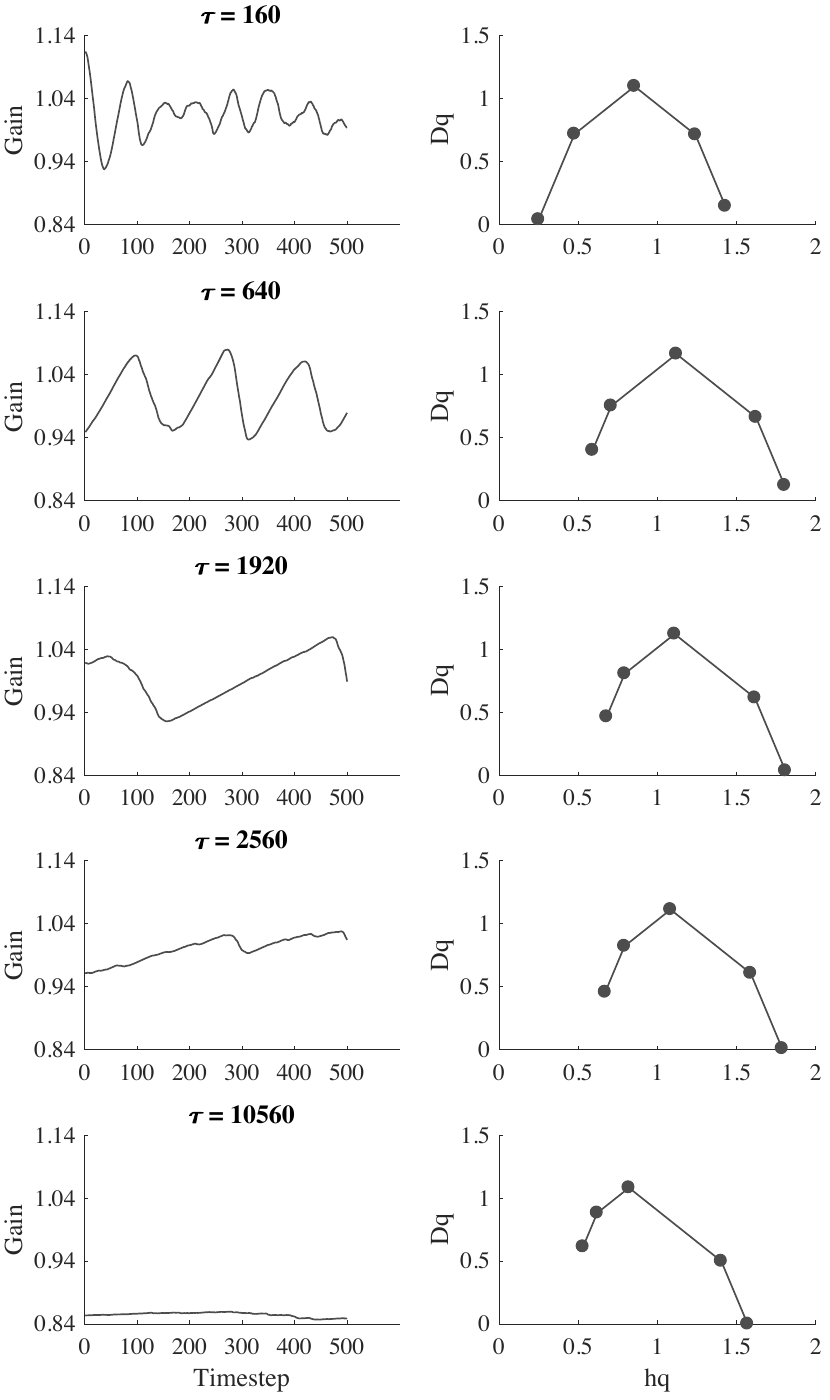}
  \caption{First $t = 500$ time steps of neuronal gain time-series (left column) and multifractal spectra of corresponding time-series ($50,000$ timesteps; right column) for select $\tau$ values ranging $\tau = 160$ to $10,560$. Gain time-series with lower characteristic time of neuronal gain recovery ($\tau$) demonstrate higher frequency fluctuations, as compared to gain time-series for higher $\tau$. Fast timescale fluctuations are reflected in multifractal spectra with extended left tails. As $\tau$ increases, slow-frequency and low-amplitude timescale fluctuations of neuronal gain become more pronounced, such that left multispectra tails are truncated compared to right tails.}
  \label{fig:2}
 \end{center}
\end{figure}

MFDFA applied an additional parameter to the DFA analysis, such that minimum and maximum $q$-orders were selected to examine the influence of segments with large and small fluctuations on the degree of neuronal gain fractal scaling. The $q$-orders ranged from $q_{min} = -5$ to $q_{max} = 5$ with a step size of $q_{step} = 2$~\cite{Ihlen2012}. Neuronal gain time-series exhibited multifractality, with average minimum Hurst exponents $M = 0.56 \pm 0.12$ and maximum Hurst exponents $M = 1.70 \pm 0.12$ having an average multifractal spectra width $M =  1.06 \pm 0.33$. Multifractal spectra width was negatively correlated with higher $\tau$, $r = -0.66$, $p = 0.02$, such that $1/f$ scaling stabilized with greater neuronal gain $M = 1.06 \pm 0.33$; see Figure~\ref{fig:2}.

In terms of overall multifractal spectra tail length, multifractal spectra exhibited elongated right tails ($M_{left-right} = 0.33 \pm 0.23$). Differences between left and right tail lengths demonstrated extended right tails for neuronal gain plasticity ranging between $\tau = 320$ and $\tau = 12,320$ ($M_{left-right} = 0.39 \pm 0.18$), while extended left tails were observed for gain plasticity between $\tau = 160$ and $\tau = 240$ ($M_{left-right} = -0.08 \pm 0.04$). Neuronal plasticity was positively associated with left tail values ($r = 0.75$, $p = 0.001$, $M = 0.43 \pm 0.17$) and negatively associated with differences between left and right tail values ($r = -0.73$, $p = 0.02$, $M = 0.33 \pm 0.23$). Elongated right tail lengths indicate that networks with higher levels of plasticity are relatively insensitive to local fluctuations with large magnitudes. 

\section{Discussion}
The present study evaluated the fractal dynamics of fully connected networks of stochastic integrate-and-fire spiking neurons in order to examine neuronal gain plasticity as a factor that may drive the brain to reach and maintain critical states. Our findings demonstrate the appropriateness of utilizing monofractal and multifractal detrended fluctuation analyses to assess critical regimes. 

We verify fractal scaling as an indicator of criticality by examining its relationship to the average branching ratio of active neurons and average neuronal gain. In addition, we demonstrate that fractal scaling is reduced within networks exhibiting more extreme (sub/supercritical systems) versus intermediate values of average neuronal gain. Networks poised at quasi-critical states must be stable enough to maintain organization but unstable enough to be adaptable. 

Neuronal gain time-series exhibit multifractal structure, indicating that multiple scaling relationships occur across each network’s signal (Figure ~\ref{fig:2}). This finding is compatible with natural temporal and spatial variation in scale-invariant structure of biomedical signals~\cite{Lopes2016}. Given that multifractal structures reflect the relative influence of various scales within a system, this finding indicates that a broad range of scales exert a meaningful effect on the network, especially for larger average gains (lower $\tau$ values). 

Multifractal spectra shifted from elongated left and shortened right tails with faster neuronal gain recovery after spiking (smaller $\tau$), to shortened left tails and elongated right tails with slower recovery time (larger $\tau$; Figure~\ref{fig:2}). Specifically, networks with small $\tau$ values exhibited local fluctuations with relatively high-amplitude fluctuations, such that multifractal spectra were relatively sensitive to local fluctuations with larger amplitudes than to less-pronounced fluctuations. Along these lines, networks with greater $\tau$ exhibited low-frequency amplitude fluctuations, such that multifractal spectra were relatively sensitive to smaller fluctuations, as compared to larger amplitude fluctuations. 

In summary, the present study identifies fractal scaling in neuronal networks as a viable measure for identifying critical states. In addition, the present study indicates that neuronal gain plasticity may play a significant role in modulating system criticality. Future work will address whether the results presented here are consistent for smaller or larger network sizes. Additionally, fractal analyses may also be fruitfully applied to time-series of other signals related to neuronal activity, including empirical recordings of \textit{in vivo} and \textit{in vitro} neuronal networks. 

\section*{Acknowledgments}
This article was produced as part of the activities of FAPESP  Research, Innovation and Dissemination Center for Neuromathematics (grant \#2013/07699-0, S.Paulo Research Foundation). AAC also thanks grants $\#$2016/00430-3 and $\#$2016/20945-8 São Paulo Research Foundation (FAPESP).

\bibliographystyle{splncs}
\bibliography{bibsuper}

\end{document}